# A New Malware Detection System Using a High Performance-ELM method

Shahab Shamshirband[1], Anthony T. Chronopoulos[2,3]

[1]Department of Computer and Information Science, Norwegian University of Science and Technology Norway
[2]Department of Computer Science, University of Texas at San Antonio, USA
[3] (Visiting Faculty) Dept of Computer Engineering & Informatics, University of Patras, Greece

*Abstract*

A vital element of a cyberspace infrastructure is cybersecurity. Many protocols proposed for security issues, which leads to anomalies that affect the related infrastructure of cyberspace. Machine learning (ML) methods used to mitigate anomalies behavior in mobile devices. This paper aims to apply a High Performance Extreme Learning Machine (HP-ELM) to detect possible anomalies in two malware datasets. Two widely used datasets (the CTU-13 and Malware) are used to test the effectiveness of HP-ELM. Extensive comparisons are carried out in order to validate the effectiveness of the HP-ELM learning method. The experiment results demonstrate that the HP-ELM was the highest accuracy of performance of 0.9592 for the top 3 features with one activation function.

## 1. Introduction

The popularity of smartphones is growing recently. According to the Gartner, the total number of smartphone's user will reach six billion by 2020 ( according to Ericsson [1] ), also more than 40 million attacks by malicious mobile malware ( according to Kaspersky Labs [2] ), and a data breach will exceed $150 million by 2020 (according to Juniper research [3]). Smartphone devices are affected by malicious malware which installs their modules in the system directory and attempts to subvert the typical system's behavior [4].

To cope with these problems, researchers and security analysts conducted a study for practical or scientific purposes to establish reliable applications for mobile devices. In [5], Russon discovered various types of hidden malware in more than 104 Google play applications which downloaded over 3.2 million times. It causes numerous attacks to user's mobile devices to affect the CPU loads for the system[5].

Several security protocols applied for malware detection. Such systems can be either anomaly-based or behavior-based. The former relies on a predefined pattern. This type of Intrusion Detection System (IDS) is an efficient approach for identifying sweeps and probes of network hardware and hints early warnings of potential intrusions before firing attacks such as Telnet. Such type systems depend on receiving regular signature updates (i.e., the extent of the signature database). The dramatic influence of DDoS attacks by Mirai bot net and its variants highlights the risks for IoT devices [6]. An end to end security scheme called Datagram Transport Layer Security (DTLS) protocol utilizes an encryption technique [7]. Although DTLS mitigate the specific type of attacks, it fails to identify all type of big data based anomalous behaviors. The significant drawback is that they are attack-specific [8]. Therefore, it is essential to develop methodologies and procedures to measure the uncertainty of IoT devices and its potential capacity to make a smart decision in order to increase the efficiency of security and privacy issue in IoT environments. It is also essential to provide smart recommendation in terms of fraud or vulnerability detection using leaning algorithms.

Soft computing techniques such as neuro fuzzy have been proposed to mitigate the security issue in IoT based malware detection [9]. Neuro-Fuzzy classifiers trained from static data and data that applications generate during their execution. Neuro-Fuzzy has a potential in malware detection based on collected statistics and derived fuzzy rules ([10], [11], [12]). Moreover, several learning algorithms have been proposed to identify the malware codes and their behaviors and identify the specific type of threats ([13],[14]). The main drawback of soft computing technique such as neuro fuzzy is that the fuzzy rules are randomly generated to tune the weights of the neural network and the number of hidden layers can increase depending on the type of application. However, there is no way to speed up the procedure of tuning. Thus, ELM can adjust the activation functions in the Single Hidden Layer Feed-forward Neural Networks (SLFNs) ([15] , [16] , [17]) and [18].

In this paper, we utilized a fast convergent reinforcement solution named High Performance Extreme Learning Machine (HP-ELM) to help the learning phase of the method to call the parameters of the hidden neurons that created randomly, which is independent of the training data [19]. Furthermore, such a method is used to test various scales of data sets, different structure selection options, and regularization methods. In our study, HP-ELM is applied to classify the malware in two datasets.

In our study, we deal with the following research questions: 1. "What are the current data analytic techniques that are being used to extract meaningful IoT based malware devices values?", 2. "How to maintain malware influence on IoT?" and 3. "What is the effect of the proposed security and privacy preserving framework in terms of scalability in the big data platform?". Our work contributes to forensic malware behaviour. Previous results and datasets of mobile malware applications used in this study ([20] ,[21]).

The contribution of the paper is to propose an IDS method to recognize IoT malware as follows.

- We applied a feature selection method in two malware datasets.
- We developed a malware detection system for IoT environment based on HP-ELM classifier.
- We tested the efficiency of the proposed system using two benchmark datasets: CTU-13 dataset [20] and DyHAP malware dataset [21] and we compared the performance of HP-ELM with and without feature selection.

The manuscript organized as follows. The previous works discussed in Section 2. Section 3 presents the HP-ELM detection method and Section 4 discusses data preparation. Section 5 presents scenarios and the setting of the HP-ELM parameters. Section 6 presents the simulation setup and evaluation metrics — experiments presented in Section 7. Section 8 concludes the paper and presents future directions.

## 2. Previous work

This section presents technical papers which use the security framework for malware based IoT [22,23]. In [22], an application program is tested in a scrutinizing manner without the implementation of the actual application (i.e., a reverse engineer process) while in [23], a program investigates the behavior of the running processes by executing the application. On the one hand, a static type malware program requires low memory resources, minimal CPU processes and the analysis process is fast, on the other hand, a dynamic one could be used to detect unknown changes and malware existence [24]. A machine learning based malware detection proposed in [25] which used a private and a public dataset. Finally, they validated their solutions in various cases. In our paper, we also evaluated the proposed methods in a private (Andoird Malware) and a public (CTU-13) dataset using a different type of HP-ELM parameter setting experiments.

Authors in [24] use a crowdsourcing system in order to attain the flows of the application's behavior. Many studies have been carried out on mobile phone malware based on a single operating system or a comparative study between two operating systems. Authors in [22] present a multilevel and behavior-based Android malware detection using 125 existing malware families and report 96% detection of malware. However, this approach applies system calls which contains less semantic information and is not able to detect malicious behavior accurately. Recently, authors in [26] captured system calls and "binder transactions" which runs the runtime behavior signature. This method presents a new malware variant detection client-server system which jointly covers the logic structures and the runtime behaviors of mobile applications for Android devices. However, this approach depends on the network status, graph mining; it impacts network performance, which directly influences the computation complexity. Therefore, it is not suitable for real-time detection.

In [27] authors present a lightweight detection system to identify malicious behaviors from mobile devices. Statistical Markov chain models applied to build the application behavior in the form of a feature vector and the random forest is adopted to classify the application behavior. The results indicate an accuracy of 96%.

Mirai attempts to categorize remarkable DDoS attacks affecting high profile targets [6]. It is a wake-up call for the control in IoT devices and analyzes the risk of increasingly DDoS attacks. It diminishes the administrative credentials of IoT devices using brute force, relying on a small dictionary of a possible username and password pairs [28]. To cope with such issues, we need resource efficiency hybrid IDSs as novel security solutions that can efficiently protect devices against DDoS, considering the insufficient resources available in the IoT environment.

## 3. HP-ELM detection method

The main drawback of soft computing technique is that the fuzzy rules are randomly generated to tune the weights of the neural network and the number of hidden layers can increase depending on the type of application. However, there is no way to speed up the procedure of tuning. Thus, ELM able adjust the activation functions in the single hidden layer feed-forward neural networks (SLFNs) ([15] , [16] , [17]) and [18].

This section presents the HP-ELM mobile network architecture applied on two malware datasets Android Malware [21] and CTU-13 dataset [20]. As mentioned in Figure 1, ELM is fast training method for SLFN networks[29].

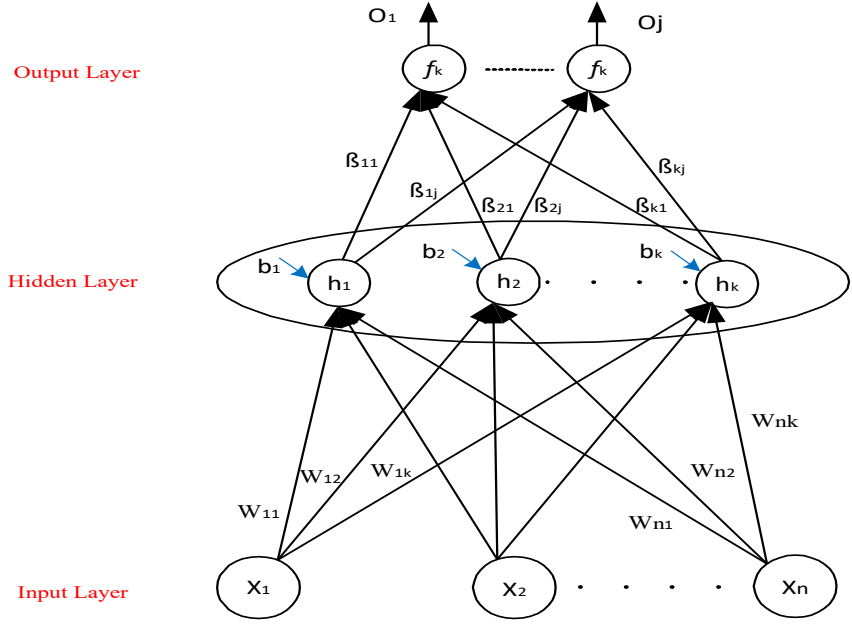

Figure 1. Computing the output of an SLFN (ELM) model

Figure 1 shows the computing of output of an ELM model. It includes three layers of neurons. There is no computation in layer one (input). The input layer weights $\omega$ and biases $b$ are set randomly and never adjusted (random distribution of the weight). The output layer is linear and there is no transformation function and bias for the output-layer. Therefore, the computing time is very fast. The word " Single" in "SLFN" is because there is only one layer of non-linear neurons (hidden layer). The main advantage of ELM is to produce weakly connected hidden layer features, because the input layer weights randomly generated, and it improves the generalization properties of the solution of a linear output layer [19].

The ELM method described as follows. We consider a set of $N$ distinct training samples $(x_i, t_i)$, $i \in [1, N]$ with $x_i \in R^c$ and $t_i \in R^c$. A SLFN with $L$ hidden neurons has the following output equations:

$\sum_{j=1}^{L} \beta_j \emptyset(w_j x_i + b_j), \quad i \in [1, N],$ (1)

where $\emptyset$ is the activation function (a sigmoid function is a common choice, but other activation functions are possible including linear [17], [18] and [29]), $w_j = [w_{j1}, w_{j2}, w_{jn,}]^T$ is the weight vector that connects the n input nodes to the jth hidden node, $b_i$ are the biases values of the jth node.

A hidden node and $\beta_j = [\beta_{j1}, \beta_{j2}, \beta_{jm,}]^T$ is the set of values of the output weights that connects the jth hidden node with m output nodes. The relation between inputs $x_i$ of the network, target outputs $t_i$ and estimated outputs $y_i$ is:

$y_i = \sum_{i=1}^{L} \beta_j \emptyset(w_j x_i + b_j) = t_i + \epsilon_i, \quad i \in [1, N],$ (2)

where $\emptyset$ is the activation function and $w_j$ is the weight vector that connects the n input nodes to the jth hidden node, $b_i$ the biases values of the jth node. The noise ($\epsilon$) includes both random noise and dependency on variables not presented in the inputs $X$.

For $N$ samples, the $N$ equations can be represented as H$\beta$=T, where

$$H = (w_1, w_2, \ldots, w_k, b_1, b_2, \ldots, b_k, x_1, x_2, \ldots, x_N) \quad (3)$$

$$= \begin{pmatrix} \emptyset(w_1.x_1 + b_1) & \cdots & \emptyset(w_k.x_1 + b_k) \\ \vdots & \ddots & \vdots \\ \emptyset(w_1.x_N + b_1) & \cdots & \emptyset(w_k.x_N + b_k) \end{pmatrix}_{N*k},$$

$$\beta = [\beta_1^T \ldots \beta_k^T]_{k*m}^T \quad (4)$$

$$T = [y_1^T \ldots y_k^T]_{N*m}^T \quad (5)$$

The output weight matrix $\beta$ is found by solving the least square problem:

$$\dot{\beta} = min_\beta \|H\beta - T\| = H^\dagger T, \quad (6)$$

$$\dot{\beta} = (H^T H)^{-1} H^T T \quad (7)$$

where $H^\dagger$ is the MP pseudo inverse of the hidden layer output matrix $H$.

This paper used HP-ELM toolbox which supports multi-class, weighted multi-class and multi-label as a classifier [19]. Section 4 describes how HP-ELM adapted to the two malware scenarios.

## 4. Dataset preparation

In this study, the HP-ELM methods evaluated on two scenarios. The first scenario uses the Mobile Malware dataset and the second scenario uses the CTU-13 dataset. We will describe the method of data collection for these two scenarios next.

**4.1. Mobile Malware dataset (Scenario 1):**

This scenario consists of two types of applications such as benign and malware. Twenty normal apk file downloaded from Google Play. These benign files installed on an Android-based operating system Jelly Bean version 4.3 which runs on mobile devices. After installation, the network traffic of running apps captured in a real time network environment in order to authenticate the behavior of apps. In the case of malware, the experiment utilizes Malgenome [30] as the malware dataset. It contains 1260 samples which consist of 49 malware families used in the previous study [21]. The identification includes several malware types, such as a botnet, root exploit, and Trojan. Authors selected seven connection-based features on the Information Gain (IG) [31] algorithm to analyze it because of its practical measuring features [32]. IG shows their strength in accuracy enhancement, the capability of generalization and short execution time. It determines how the training sets separated according to the target classification [33]. In Scenario 1, the higher gain ratio indicates the feature's relevance in a classification model for a machine learning classifier. Therefore, the features are maximum_frame, frame_STD, count_ACK, Minimum_Dest_Port, Average_Frame, and Average_source_port [21] . Figure 2 shows the methodology of collection of mobile malware dataset in Scenario 1.

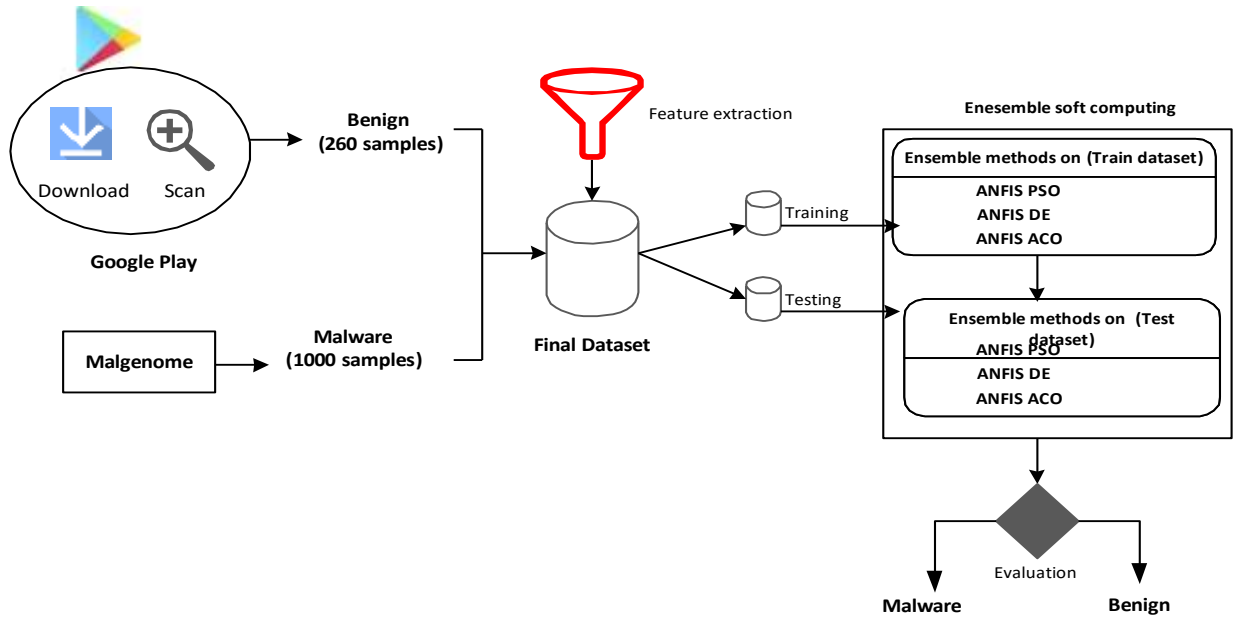

Figure 2: The methodology of collection of Mobile Malware data

## 4.2. CTU-13 dataset (Scenario 2):

This dataset consists of thirteen captures (called Scenarios) of different real botnet samples [20]. The characteristics of the scenarios and the features (pcap files) captured in terms of tcpdump. The pcap files are converted to netflow file standard using the argus software suite in two steps. The first step converts the pcap files to a bidirectional argus binary storage. The second step converts the argus bin to Netflow. The outcome of these steps is the final netflow file [34]. The next step is to assign the label to the netflow data. The background label assigns the normal label to the traffic which matches a specific filter. Then the botnet label is assigned to the traffic that comes from or to any of the known infected IP address [20]. Figure 3 indicates the methodology of collection of CTU-13 and how the HP-ELM classifies it.

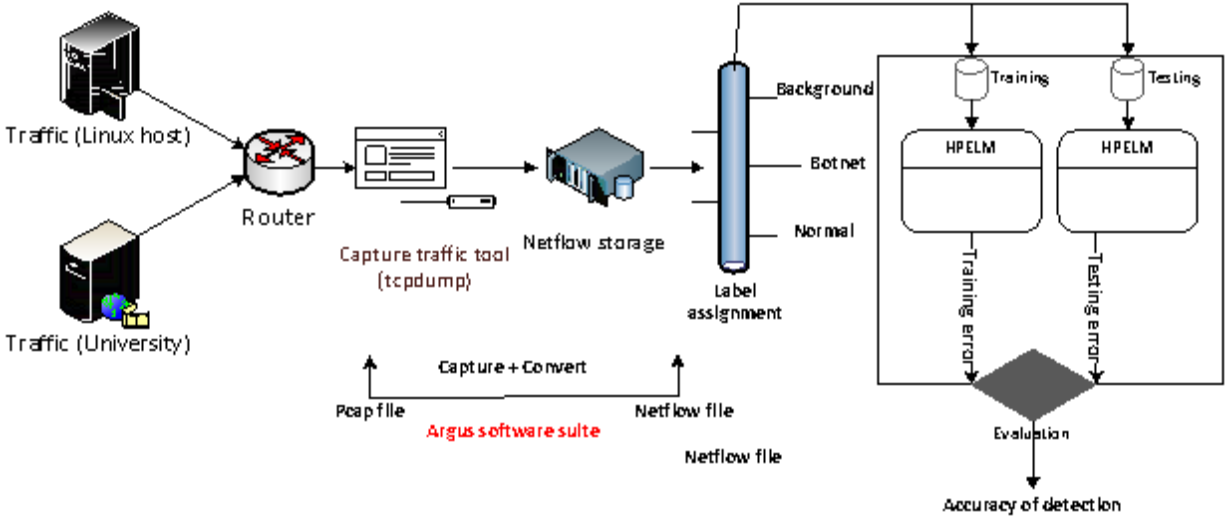

Figure 3: The methodology of collection of CTU-13

In this research work, the labeled dataset splits into training (70%) and testing data (30%) [35] to evaluate with the HP-ELM algorithm. In the next section, we propose the procedure of applying the method to the datasets.

## 5. Proposed method

The proposed method includes a few modules. In detail, the first module (Reading) is used to read the malware datasets continuously and transfer them to feature selection modules. Then, the second module (Filtering) is responsible for normalizing the data and applying the feature selection algorithm to select the essential features for the proposed algorithm. After that, the third module (Splitting) is responsible for splitting the data into 70% training and 30% testing. Then, the fourth module (HP-ELM) is adapted to tune its parameters to predict the targets. Finally, the fifth module (Evaluation) is responsible for predicting the actual values in the training and testing phase.

### 5.1. Feature selection

Before applying the HP-ELM method, we utilized the F-Score [36] and Fisher Score [37] in order to select the most valuable features. Then, the HP-ELM builds to new data. The extracted features from CTU-13 dataset are: {DstAddr, sport, Proto, SrcAddr, Dport, Dur, State, sTos, dTos, TotBytes, SrcBytes and TotPkts} [38] . More details about the features of CTU-13 mentioned in [20].

### 5.2. HP-ELM parameter setting (activation functions, number of neurons, etc)

We applied six types of activation functions in the layer of HP-ELM such as linear, rbf-linf, rbf-l1, rbf-l1, tanh, and sigmoid. A total of 2000 of neurons applied in the layer of HP-ELM. We investigated HP-ELM with feature selection (Top 3, and top 5 features) and without feature selection (all features). We also applied different strategies for activation functions in the layer of HP-ELM (one, two, three and four activation functions) in both datasets. Finally, the state of the best answer recorded in the tables below.

## 6. Performance Evaluation

The following subsections describe in detail the pursued datasets, evaluation metrics, scenarios and their settings, and the results for the applied scenarios.

### 6.1. Simulation Setup

The methods tested on a system with Intel Core i7CPUand 8-GB RAM. Two benchmark malware datasets CTU-13 dataset (Scenario 5) [20] and DyHAP malware dataset [21] have been used to evaluate the performance of HP-ELM.

### 6.2. Metrics

This research used the accuracy ratio of Equation 8 to evaluate the performance of the method. In general, the accuracy is the number of applications which the classifier correctly detects, divided by the total number of malicious and legitimate applications. The accuracy is between 0 and 1, i.e. $0 \leq Accuracy \leq 1$. We also have

$$Accuracy\ Ratio = \frac{TP+TN}{TP+FN+TN+FP} \quad (8)$$

where the used parameters denote the following numbers. TN is the accurately classified benign instances; TP is the malicious applications that are appropriately identified, FP is the wrongly classified benign instances as malware applications; and, FN is the malware instances wrongly classified as a legitimate application.

## 7. Results and discussion

We evaluated the results using various layers with/without the feature selection.

### 7.1. Evaluate HP-ELM without applying feature selection in android malware

This subsection compared the performance of the HP-ELM in the presence of various active functions for two datasets without using feature selections and integrated all features. A total number of 13 features are available in CTU-13 dataset, and the number of total input features for Android Malware dataset is 6. Table 1 shows the accuracy rate of HP-ELM in training (70%) and testing (30%). Table 1 shows that the HP-ELM reach a high accuracy of 0.9696 in training and 0.9672 in testing with 2000 neurons of Rbf-linf activation function without applying the feature selection. The result indicates that the accuracy of HP-ELM increased with two activation functions (Rbf_linf(1000), Sigmoid(1000)). Finally, the accuracy of HP-ELM with three activation functions Sigmoid (400), Tanh (1000), Rbf_linf(600), and a total number of 2000 neurons reach 0.9679 in training and 0.9661 in testing.

**TABLE 1**
The accuracy rate of the HP-ELM method with different active functions and without feature selection for training and test data of **Malware** datasets

| Priority of Feature | One Activation Function | Accuracy Ratio (%) | |
|---|---|---|---|
| | | Train Data | Test Data |
| | | 70% | 30% |
| All Features [0 5 6 4 2 3 1] | Linear (2000) | 0.7087 | 0.7095 |
| | Rbf_l1 (2000) | 0.9694 | 0.9669 |
| | Tanh (2000) | 0.9654 | 0.9636 |
| | Sigmoid (2000) | 0.9657 | 0.9637 |
| | Rbf_l2 (2000) | 0.9678 | 0.9653 |
| | Rbf_linf(2000) | 0.9696 | 0.9672 |
| | **Two Activation Function** | Train Data | Test Data |
| All Features [0 5 6 4 2 3 1] | Sigmoid (1000), Linear (1000), | 0.9589 | 0.9581 |
| | Rbf_l2(1000), Linear (1000) | 0.9613 | 0.9594 |
| | Rbf_l1 (1000), Sigmoid (1000) | 0.9680 | 0.9659 |
| | Tanh (1000), Rbf_l1(1000) | 0.9679 | 0.9662 |
| | Sigmoid (1000), Tanh (1000) | 0.9664 | 0.9638 |
| | Rbf_linf(1000), Sigmoid(1000) | 0.9689 | 0.9673 |
| | **Three Activation Function** | Train Data | Test Data |
| All Features [0 5 6 4 2 3 1] | Tanh (750), Rbf_linf(250), Linear (1000) | 0.9610 | 0.9593 |
| | Tanh (1000), Rbf_linf(550), Linear (450) | 0.9648 | 0.9635 |
| | Sigmoid (400), Tanh (1000), Rbf_linf(600), | 0.9679 | 0.9661 |

Figure 4 (a, b) indicates the accuracy of HP-ELM with and without feature selection with one, two and three activation function for android malware dataset. The x-axis represents the activation function's name. For example, the order of one activation function with all features are linear, Rbf_l1, Tanh, Sigmoid, Rbf_l2 and Rbf_linf which are mapped to the tag point (0.0, 0.5,1.0,1.5,2.0,2.5,3.0,3.5,4.0).

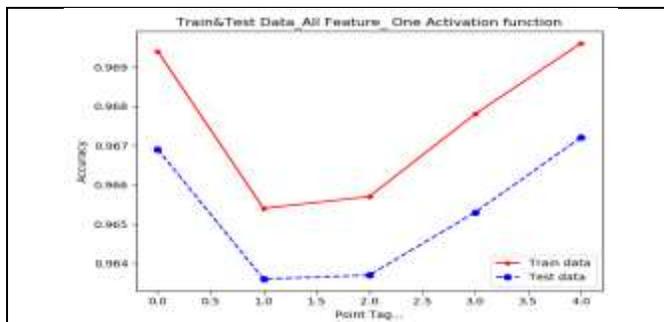

Fig 4 (a). Accuracy of HP-ELM without feature selection with one activation function for android malware dataset

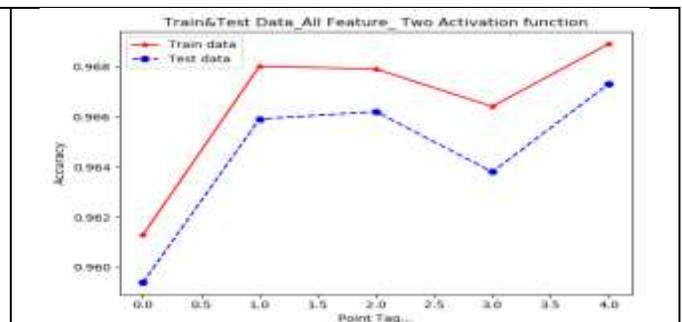

Fig 4 (b). Accuracy of HP-ELM without feature selection with Two activation function for android malware dataset

## 7.2. Evaluate HP-ELM by applying feature selection (Top 3 features) on android malware

This subsection compares the performance of the HP-ELM in the presence of various active functions for the android malware datasets with three higher priority features using F-Score selection policy. The goal of this scenario is to analyze the accuracy rate of the HP-ELM method in the presence of different activation functions using various selected features. In Table 2, we have numerically tested this, by applying the top 3 features, the accuracy of HP-ELM in terms of one activation function Rbf_l1(2000) reach a high value of 0.9056 in training and 0.9017 in testing. On the other hand, a combination of two activation functions such as Tanh (1000), Rbf_l1(1000) Hp-elm gives a better accuracy ratio of 0.9018 in testing. Finally, the high accuracy reached with three activation function Sigmoid (400), Tanh (1000), Rbf_linf(600) in testing 0.8998. It can be considered the best strategy rule of a two-activation function.

**TABLE 2**
The accuracy rate of the HP-ELM method with different active functions and different priority of feature (with feature selection) for training and test data of **Malware** datasets (Top 3 feature)

| Priority of Feature | One Activation Function | Accuracy Ratio (%) | |
|---|---|---|---|
| | | Train Data | Test Data |
| | | 70% | 30% |
| Top 3 Features [0 5 6] | Linear (2000) | 0.6928 | 0.6923 |
| | Sigmoid (2000) | 0.8941 | 0.8918 |
| | Tanh (2000) | 0.9017 | 0.8975 |
| | Rbf_l1 (2000) | 0.9056 | 0.9017 |
| | Rbf_l2 (2000) | 0.8873 | 0.8853 |
| | Rbf_linf (2000) | 0.9023 | 0.8978 |
| | Two Activation Function | Train Data | Test Data |
| Top 3 Features [0 5 6] | Sigmoid (1000), Linear (1000), | 0.8880 | 0.8846 |
| | Sigmoid (1000), Tanh (1000) | 0.9007 | 0.8968 |
| | Tanh (1000), Rbf_l1(1000) | 0.9058 | 0.9016 |
| | Rbf_l1 (1000), Sigmoid (1000) | 0.9046 | 0.9018 |
| | Rbf_l2(1000), Linear (1000) | 0.8843 | 0.8824 |
| | Rbf_linf(1000), Sigmoid(1000) | 0.8995 | 0.8946 |
| | Three Activation Function | Train Data | Test Data |
| Top 3 Features [0 5 6] | Sigmoid (400), Tanh (1000), Rbf_linf(600), | 0.9037 | 0.8998 |
| | Tanh (1000), Rbf_linf(550),Linear (450) | 0.8993 | 0.8963 |
| | Tanh (750), Rbf_linf(250), Linear (1000) | 0.8909 | 0.8883 |

Figure 5a, and 5b indicate the accuracy of HP-ELM in android malware scenario.

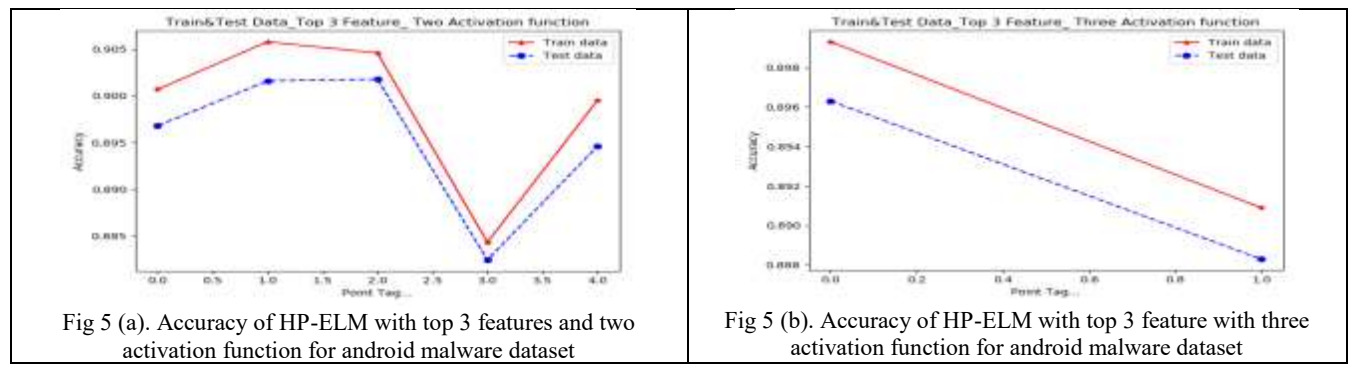

Fig 5 (a). Accuracy of HP-ELM with top 3 features and two activation function for android malware dataset

Fig 5 (b). Accuracy of HP-ELM with top 3 feature with three activation function for android malware dataset

## 7.3. Evaluate HP-ELM with applying feature selection (Top 5 features) on android malware

This subsection compares the performance of the HP-ELM in the presence of various active functions for the same layer for android malware datasets with five higher priority features using F-Score selection policy. The goal of this scenario is to analyze the accuracy rate of the HP-ELM method in the presence of different activation functions using various selected features. In Table 3, we have numerically tested this, by applying top 5 features, the accuracy of HP-ELM in terms of one activation function Rbf_linf (2000) reach to the high value of 0.9675 in training and 0.9656 in testing. However, the testing result of Rbf_l1 (2000) is higher than Rbf_linf (2000), which is 0.9702.

**TABLE 3**
The accuracy rate of the HP-ELM method with different active functions and different priority of feature (with feature selection) for training and test data of **Malware** datasets (Top 5 features)

| Priority of Feature | One Activation Function | Accuracy Ratio (%) | |
|---|---|---|---|
| | | Train Data | Test Data |
| | | 70% | 30% |
| Top 5 Features [0 5 6 4 2] | Linear (2000) | 0.7009 | 0.7019 |
| | Sigmoid (2000) | 0.9663 | 0.9642 |
| | Tanh (2000) | 0.9651 | 0.9644 |
| | Rbf_l1 (2000) | 0.9713 | 0.9702 |
| | Rbf_l2 (2000) | 0.9633 | 0.9622 |
| | Rbf_linf (2000) | 0.9675 | 0.9656 |
| | **Two Activation Function** | **Train Data** | **Test Data** |
| Top 5 Features [0 5 6 4 2] | Sigmoid (1000), Linear (1000), | 0.9583 | 0.9578 |
| | Sigmoid (1000), Tanh (1000) | 0.9660 | 0.9643 |
| | Tanh (1000), Rbf_l1(1000) | 0.9693 | 0.9683 |
| | Rbf_l1 (1000), Sigmoid (1000) | 0.9707 | 0.9689 |
| | Rbf_l2(1000), Linear (1000) | 0.9608 | 0.9610 |
| | Rbf_linf(1000), Sigmoid(1000) | 0.9673 | 0.9654 |
| | **Three Activation Function** | **Train Data** | **Test Data** |
| Top 5 Features [0 5 6 4 2] | Sigmoid (400), Tanh (1000), Rbf_linf(600) | 0.9668 | 0.9661 |
| | Tanh (1000), Rbf_linf(550), Linear (450) | 0.9635 | 0.9616 |
| | Tanh (750), Rbf_linf(250), Linear (1000) | 0.9583 | 0.9572 |

Although, the result of a combination of two activation functions Rbf_linf (1000), Sigmoid (1000), Hp-elm gives a better accuracy ratio of 0.9673 in training and 0.9654 in testing, but the combination of two activation functions of Rbf_l1 (1000), Sigmoid (1000) reaches a high accuracy in testing phase which is equal to 0.9689. The combination of three activation functions such as Sigmoid (400), Tanh (1000), Rbf_linf(600), gives the accuracy of 0.9668 in training and 0.9661 in testing which is not very high in comparison with the cases of two and one activation functions.

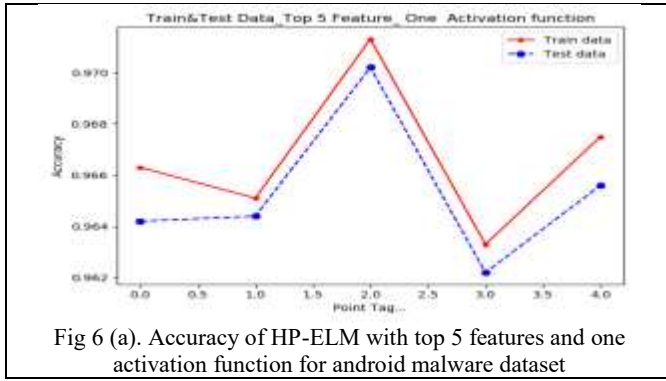

Fig 6 (a). Accuracy of HP-ELM with top 5 features and one activation function for android malware dataset

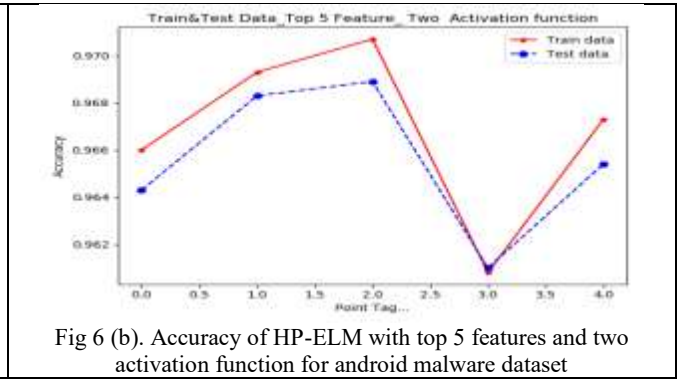

Fig 6 (b). Accuracy of HP-ELM with top 5 features and two activation function for android malware dataset

In Figure 6, we confirm that when the number of activation function increases, the accuracy will also increase, and its value will remain constant after 1000 neurons in datasets. By increasing the number of activation (two) functions, the highest accuracy reached in the testing phase.

**TABLE 4**
The accuracy rate of the HP-ELM method without feature selection and different active functions in training and test data of **CTU-13** datasets

| Priority of Feature | Activation Function | Accuracy Ratio (%) CTU-13 Dataset (Scenario 5) | |
|---|---|---|---|
| | | Train Data 70% | Test Data 30% |
| | One activation function | | |
| Features [3 7 1 4 6 2 5 0 12 9 11 10 8] | Linear (2000) | 0.3095 | 0.3144 |
| | Rbf_l1 (2000) | 0.9599 | 0.9563 |
| | Tanh (2000) | 0.9635 | 0.9597 |
| | Sigmoid (2000) | 0.9637 | 0.9592 |
| | Rbf_l2 (2000) | 0.9654 | 0.9603 |
| | Rbf_linf(2000) | 0.9650 | 0.9625 |
| | Two activation function | | |
| Features [3 7 1 4 6 2 5 0 12 9 11 10 8] | Tanh (1000), Sigm(1000) | 0.9631 | 0.9596 |
| | Tanh (1000), rbf_l1 (1000) | 0.9621 | 0.9600 |
| | Tanh (1000), rbf_l2 (1000) | 0.9636 | 0.9603 |
| | Tanh (1000), rbf_linf (1000) | 0.9639 | 0.9621 |
| | Tanh (1000), linear (1000) | 0.9614 | 0.9596 |
| | Sigm(1000), rbf_l1(1000) | 0.9621 | 0.9596 |
| | Sigm(1000), rbf_l2(1000) | 0.9641 | 0.9604 |
| | Sigm(1000), rbf_linf(1000) | 0.9639 | 0.9615 |
| | Sigm(1000), linear(1000) | 0.9609 | 0.9588 |
| | rbf_l1(1000), rbf_l2(1000) | 0.9622 | 0.9598 |
| | rbf_l1(1000), rbf_linf(1000) | 0.9642 | 0.9622 |
| | rbf_l1(1000), linear(1000) | 0.9575 | 0.9558 |
| | rbf_l2(1000), rbf_linf(1000) | 0.9645 | 0.9619 |
| | rbf_l2(1000), linear(1000) | 0.9615 | 0.9592 |
| | rbf_linf(1000), linear(1000) | 0.9625 | 0.9606 |
| | Three activation function | | |
| Features [3 7 1 4 6 2 5 0 12 9 11 10 8] | Tanh(666), sigm(667), rbf_l1(667) | 0.9621 | 0.9593 |
| | Tanh(666), sigm(667), rbf_l2(667) | 0.9633 | 0.9598 |
| | Tanh(666), sigm(667), rbf_linf(667) | 0.9642 | 0.9616 |
| | Tanh(666), sigm(667), linear(667) | 0.9619 | 0.9594 |
| | Sigm(666), rbf_l1(667), rbf_l2(667) | 0.9623 | 0.9596 |
| | Sigm(666), rbf_l1(667), rbf_linf(667) | 0.9641 | 0.9616 |
| | Sigm(666), rbf_l1(667), Linear(667) | 0.9615 | 0.9595 |
| | rbf_l1(666), rbf_l12(667), rbf_linf(667) | 0.9638 | 0.9616 |
| | rbf_l1(666), rbf_l12(667), linear (667) | 0.9611 | 0.9591 |
| | rbf_l2(666), rbf_linf(667), linear(667) | 0.9626 | 0.9605 |
| | Four activation function | | |
| Features [3 7 1 4 6 2 5 0 12 9 11 10 8] | Tanh(500), sigm(500), rbf_l1(500), rbf_l2(500) | 0.9630 | 0.9603 |
| | Tanh(500), sigm(500), rbf_l1(500), rbf_linf (500) | 0.9633 | 0.9610 |
| | Tanh(500), sigm(500), rbf_l1(500), linear (500) | 0.9619 | 0.9596 |
| | sigm(500), rbf_l1(500), rbf_l2(500), rbf_linf (500) | 0.9633 | 0.9611 |
| | sigm(500), rbf_l1(500), rbf_l2(500), linear (500) | 0.9615 | 0.9590 |
| | rbf_l1(500), rbf_l2(500), rbf_linf (500) linear (500) | 0.9619 | 0.9599 |

### 7.4. Evaluate HP-ELM without applying feature selection on CTU-13

HP-ELM applied to CTU-13 in training and test cases. For example, with one activation function (Rbf_linf) with 2000 neurons reaches a high accuracy of classification of 0.9625 in testing. On the other hand, the higher accuracy with two activation functions rbf_l1(1000), rbf_linf(1000) is 0.9622 in the testing phase. By increasing the number of activation functions to three, the result does not reach as high accuracy in comparison with one and two activation functions. The highest accuracy with three activation functions is 0.9616 which is lower than 0.9625 obtained with one activation function and 0.9622 obtained with two activation function.

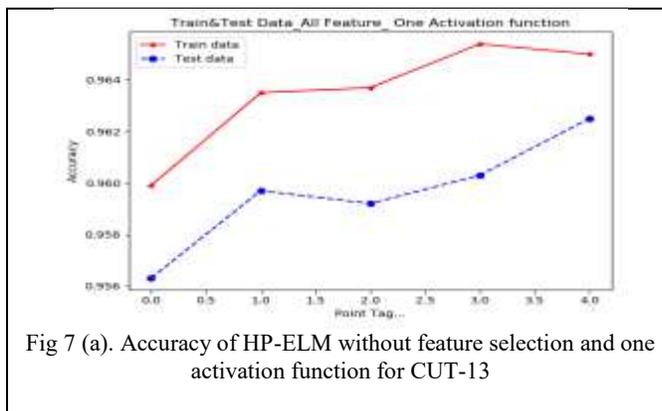

Fig 7 (a). Accuracy of HP-ELM without feature selection and one activation function for CUT-13

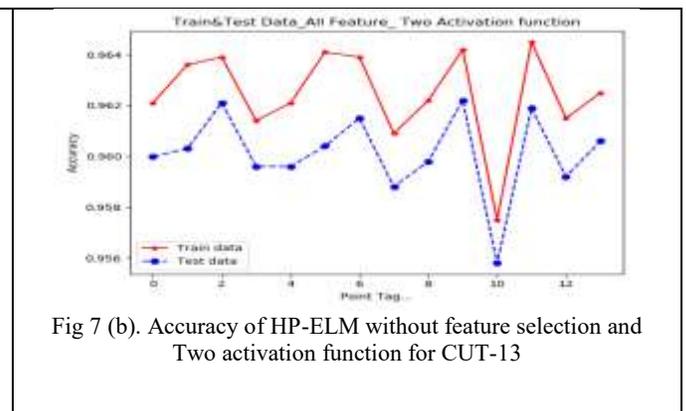

Fig 7 (b). Accuracy of HP-ELM without feature selection and Two activation function for CUT-13

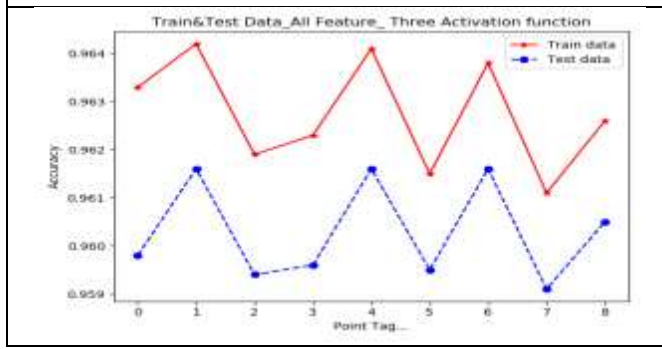

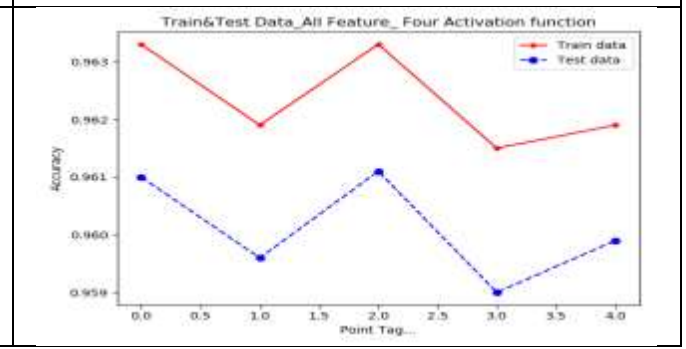

| Fig 7 (c). Accuracy of HP-ELM without feature selection and Three activation function for CUT-13 | Fig 7 (d). Accuracy of HP-ELM without feature selection and Four activation function for CUT-13 |
|---|---|

According to Figures 7 (a) ,7 (b) and 7(c) for the CTU-13 dataset, it is confirmed that more activation functions provide more complexities. It observed that the rate does not increase. Moreover, when we reach to the best number of neurons, increasing activation functions does not affect the accuracy in comparison with the case of one activation function in the layer of HP-ELM.

### 7.5. Evaluate HP-ELM with applying feature selection on CTU-13 (Top 3)

In this section, the feature selection strategy applied to CTU-13. Therefore, three high priority features are selected based on the F-score algorithm, then the selected feature allows hp-elm to evaluate the performance of the method. As in previous sections, various activation functions with a total number of 2000 neurons applied to CTU-13.

**TABLE 5**
The accuracy rate of the HP-ELM method with feature selection and different active functions in training and test data of **CTU-13** datasets (Top 3 features)

| Priority of Feature | Activation Function | Accuracy Ratio (%) | |
|---|---|---|---|
| | | CTU-13 Dataset (Scenario 5) | |
| | | Train Data | Test Data |
| | | 70% | 30% |
| | One activation function | | |
| Features [3 7 1] | Linear (2000) | 0.5473 | 0.5440 |
| | Rbf_l1 (2000) | 0.9585 | 0.9564 |
| | Tanh (2000) | 0.9593 | 0.9566 |
| | Sigmoid (2000) | 0.9585 | 0.9566 |
| | Rbf_l2 (2000) | 0.9575 | 0.9559 |
| | Rbf_linf(2000) | 0.9611 | 0.9592 |
| | Two activation function | | |
| Features [3 7 1] | Tanh (1000), Sigm(1000) | 0.9593 | 0.9570 |
| | Tanh (1000), rbf_l1 (1000) | 0.9598 | 0.9578 |
| | Tanh (1000), rbf_l2 (1000) | 0.9591 | 0.9569 |
| | Tanh (1000), rbf_linf (1000) | 0.9605 | 0.9578 |
| | Tanh (1000), linear (1000) | 0.9593 | 0.9572 |
| | Sigm(1000), rbf_l1(1000) | 0.9596 | 0.9581 |
| | Sigm(1000), rbf_l2(1000) | 0.9584 | 0.9566 |
| | Sigm(1000), rbf_linf(1000) | 0.9601 | 0.9579 |
| | Sigm(1000), linear(1000) | 0.9581 | 0.9564 |
| | rbf_l1(1000), rbf_l2(1000) | 0.9585 | 0.9563 |
| | rbf_l1(1000), rbf_linf(1000) | 0.9608 | 0.9588 |
| | rbf_l1(1000), linear (1000) | 0.9571 | 0.9551 |
| | rbf_l2(1000), rbf_linf(1000) | 0.9598 | 0.9580 |
| | rbf_l2(1000), linear (1000) | 0.9574 | 0.9559 |
| | rbf_linf(1000), linear(1000) | 0.9596 | 0.9572 |

| Features [3 7 1] | Three activation function | | |
|---|---|---|---|
| | Tanh(666), sigm(667), rbf_l1(667) | 0.9595 | 0.9577 |
| | Tanh(666), sigm(667), rbf_l2(667) | 0.9590 | 0.9571 |
| | Tanh(666), sigm(667), rbf_linf(667) | 0.9598 | 0.9576 |
| | Tanh(666), sigm(667), linear(667) | 0.9590 | 0.9572 |
| | Sigm(666), rbf_l1(667), rbf_l2(667) | 0.9587 | 0.9569 |
| | Sigm(666), rbf_l1(667), rbf_linf(667) | 0.9607 | 0.9589 |
| | Sigm(666), rbf_l1(667), Linear(667) | 0.9586 | 0.9566 |
| | rbf_l1(666), rbf_l12(667), rbf_linf(667) | 0.9603 | 0.9581 |
| | rbf_l1(666), rbf_l12(667), linear(667) | 0.9581 | 0.9557 |
| | rbf_l2(666), rbf_linf(667), linear(667) | 0.9593 | 0.9576 |
| | Four activation function | | |
| Features [3 7 1] | Tanh(500), sigm(500), rbf_l1(500), rbf_l2(500) | 0.9590 | 0.9571 |
| | Tanh(500), sigm(500), rbf_l1(500), rbf_linf(500) | 0.9605 | 0.9582 |
| | Tanh(500), sigm(500), rbf_l1(500), linear(500) | 0.9591 | 0.9573 |
| | sigm(500), rbf_l1(500), rbf_l2(500), rbf_linf(500) | 0.9596 | 0.9576 |
| | sigm(500), rbf_l1(500), rbf_l2(500), linear(500) | 0.9583 | 0.9563 |
| | rbf_l1(500), rbf_l2(500), rbf_linf(500) linear(500) | 0.9595 | 0.9574 |

Table 5 shows the best accuracy with three activation function is 0.9589 in testing phase.

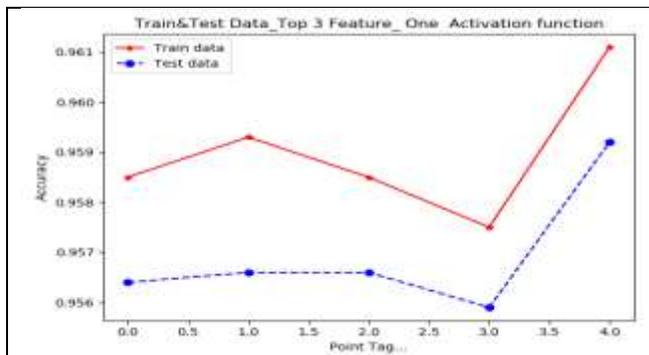

Fig 8 (a). Accuracy of HP-ELM with feature selection and one activation function for CUT-13

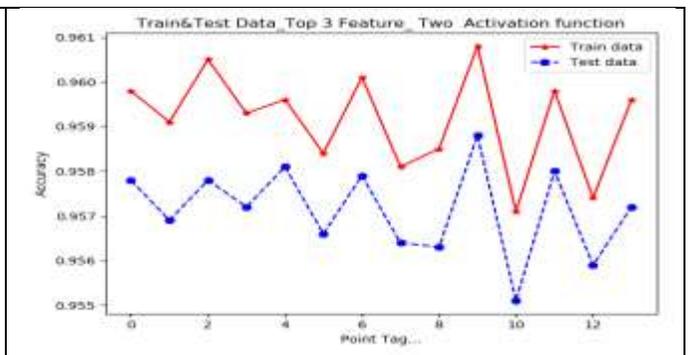

Fig 8 (b). Accuracy of HP-ELM with feature selection and two activation function for CUT-13

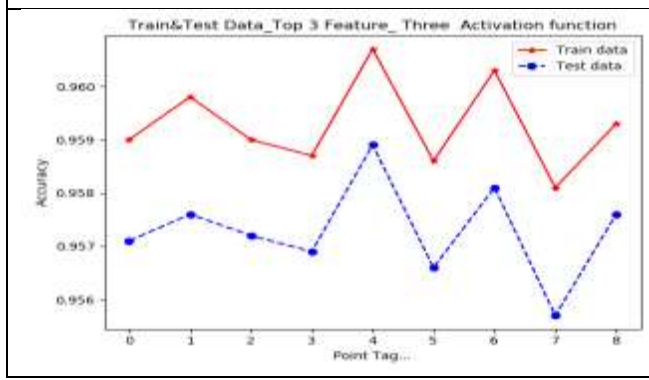

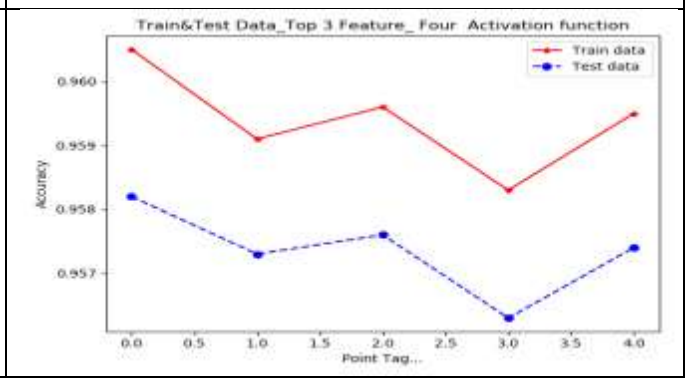

| Fig 8 (c). Accuracy of HP-ELM with feature selection and three activation function for CUT-13 | Fig 8 (d). Accuracy of HP-ELM with feature selection and four activation function for CUT-13 |
|---|---|

On the other hand, the highest accuracy with one activation function (Rbf-linf) reaches 0.9592 in the testing phase. Figure 8 (a,b,c) shows the accuracy of hp-elm with the top 3 features and different activation functions.

### 7.6. Evaluate HP-ELM with applying feature selection on CTU-13 (Top 5)

In this scenario, five high priority features are selected based on the F-score algorithm. By applying one activation function with a total number of 2000 of neurons, high accuracy of 0.9589 in Rbf_linf (2000) is reached in the testing phase as shown in Table 6.

**TABLE 6**
The accuracy rate of the HP-ELM method with feature selection and different active functions in training and test data of **CTU-13** datasets (Top 5 features)

| Priority of Feature | Activation Function | Accuracy Ratio (%) | |
|---|---|---|---|
| | | CTU-13 Dataset (Scenario 5) | |
| | | Train Data | Test Data |
| | | 70% | 30% |
| | One activation function | | |
| Features [3 7 1 4 6] | Linear (2000) | 0.4399 | 0.4399 |
| | Rbf_l1 (2000) | 0.9600 | 0.9570 |
| | Tanh (2000) | 0.9622 | 0.9564 |
| | Sigmoid (2000) | 0.9610 | 0.9578 |
| | Rbf_l2 (2000) | 0.9603 | 0.9585 |
| | Rbf_linf(2000) | 0.9616 | 0.9589 |
| | Two activation function | | |
| Features [3 7 1 4 6] | Tanh (1000), Sigm(1000) | 0.9619 | 0.9577 |
| | Tanh (1000), rbf_l1 (1000) | 0.9616 | 0.9576 |
| | Tanh (1000), rbf_l2 (1000) | 0.9612 | 0.9582 |
| | Tanh (1000), rbf_linf (1000) | 0.9616 | 0.9581 |
| | Tanh (1000), linear (1000) | 0.9606 | 0.9584 |
| | Sigm(1000), rbf_l1(1000) | 0.9615 | 0.9583 |
| | Sigm(1000), rbf_l2(1000) | 0.9607 | 0.9580 |
| | Sigm(1000), rbf_linf(1000) | 0.9613 | 0.9576 |
| | Sigm(1000), linear(1000) | 0.9606 | 0.9581 |
| | rbf_l1(1000), rbf_l2(1000) | 0.9608 | 0.9586 |
| | rbf_l1(1000), rbf_linf(1000) | 0.9612 | 0.9586 |
| | rbf_l1(1000), linear(1000) | 0.9569 | 0.9548 |
| | rbf_l2(1000), rbf_linf(1000) | 0.9611 | 0.9588 |
| | rbf_l2(1000), linear(1000) | 0.9603 | 0.9585 |
| | rbf_linf(1000), linear(1000) | 0.9605 | 0.9590 |
| | Three activation function | | |
| Features [3 7 1 4 6] | Tanh(666), sigm(667), rbf_l1(667) | 0.9615 | 0.9576 |
| | Tanh(666), sigm(667), rbf_l2(667) | 0.9613 | 0.9577 |
| | Tanh(666), sigm(667), rbf_linf(667) | 0.9616 | 0.9577 |
| | Tanh(666), sigm(667), linear(667) | 0.9609 | 0.9581 |
| | Sigm(666), rbf_l1(667), rbf_l2(667) | 0.9612 | 0.9582 |
| | Sigm(666), rbf_l1(667), rbf_linf(667) | 0.9614 | 0.9581 |
| | Sigm(666), rbf_l1(667), Linear(667) | 0.9607 | 0.9582 |
| | rbf_l1(666), rbf_l12(667), rbf_linf(667) | 0.9615 | 0.9588 |
| | rbf_l1(666), rbf_l12(667), linear(667) | 0.9604 | 0.9583 |
| | rbf_l2(666), rbf_linf(667), linear(667) | 0.9606 | 0.9586 |
| Four activation function | | | |
| Features [3 7 1 4 6] | Tanh(500), sigm(500), rbf_l1(500), rbf_l2(500) | 0.9614 | 0.9579 |
| | Tanh(500), sigm(500), rbf_l1(500), rbf_linf (500) | 0.9616 | 0.9585 |
| | Tanh(500), sigm(500), rbf_l1(500), linear (500) | 0.9613 | 0.9582 |
| | sigm(500), rbf_l1(500), rbf_l2(500), rbf_linf (500) | 0.9611 | 0.9583 |
| | sigm(500), rbf_l1(500), rbf_l2(500), linear (500) | 0.9608 | 0.9585 |
| | rbf_l1(500), rbf_l2(500), rbf_linf (500) linear (500) | 0.9606 | 0.9585 |

The second-highest accuracy with two activation function of rbf_linf(1000), linear(1000) is 0.9590 in the testing phase. On the other hand, the third highest accuracy with three activation function rbf_l1(666), rbf_l12(667), rbf_linf (667) is 0.9588 in testing phase. By increasing the number of activation functions to four, the accuracy of 0.9585 is the same in the testing phase for the activation function of 1) Tanh(500), sigm(500), rbf_l1(500), rbf_linf (500) 2) sigm(500), rbf_l1(500), rbf_l2(500), linear (500) and 3) rbf_l1(500), rbf_l2(500), rbf_linf (500) linear (500), but the training phase of Tanh(500), sigm(500), rbf_l1(500), rbf_linf (500) is better than the rest of the activation functions.

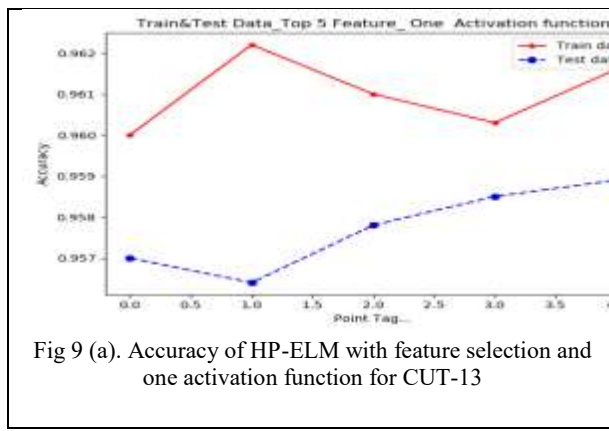

Fig 9 (a). Accuracy of HP-ELM with feature selection and one activation function for CUT-13

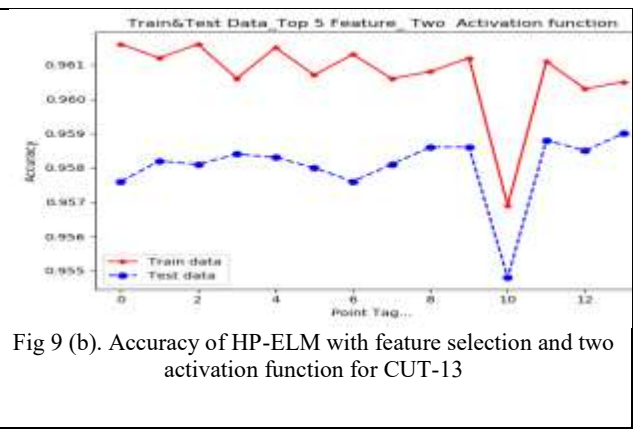

Fig 9 (b). Accuracy of HP-ELM with feature selection and two activation function for CUT-13

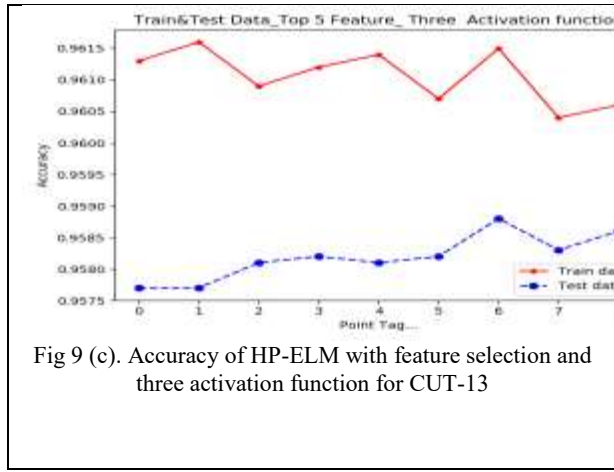

Fig 9 (c). Accuracy of HP-ELM with feature selection and three activation function for CUT-13

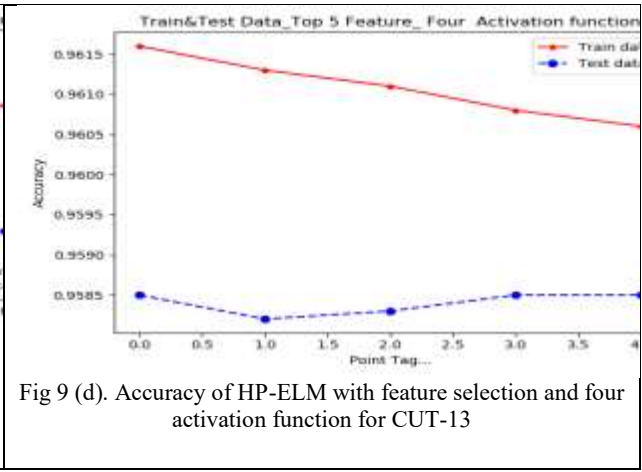

Fig 9 (d). Accuracy of HP-ELM with feature selection and four activation function for CUT-13

### 7.7. Comparison of results

For the evaluation of a proposed model, the key point is to show that the HP-ELM model is producing the smallest error in testing datasets or the highest accuracy of classification. As Table 7 shows, the accuracy values for malware dataset without feature selection are rather high with two activation function Rbf_linf (1000), and Sigmoid (1000). From Table 8 with feature selection, it is apparent that the precision is rather high in terms of two activation function of Rbf_l1 (1000) and Sigmoid (1000) compared with other metrics. Table 9 and 10 show the evaluation of the proposed method on CTU-13 with and without feature selection. The outcome of Table 9 and 10 indicate that the higher accuracy of HP-ELM in CTU-13 without feature selection is 0.9625 in the testing phase which is related to one activation function Rbf_linf(2000). On the other hand, the best accuracy of feature selection based on HP-ELM for top 3 is 0.9592 which is Rbf_linf(2000) and the higher accuracy for top 5 feature in testing is 0.9590 which is rbf_linf(1000), linear(1000).

Table 7: Accuracy of HP-ELM without feature selection in malware dataset

| Dataset | | Without feature selection | | |
|---|---|---|---|---|
| | #Activation function | | All features | |
| | | | Training | Testing |
| Malware | 1 activation function | Rbf_linf(2000) | 0.9696 | 0.9672 |
| | 2 activation function | Rbf_linf(1000), Sigmoid(1000) | 0.9689 | 0.9673 |
| | 3 activation function | Sigmoid (400), Tanh (1000), Rbf_linf(600), | 0.9679 | 0.9661 |

Table 8: Accuracy of HP-ELM with feature selection in malware dataset

| Dataset | | With feature selection | | | | | |
|---|---|---|---|---|---|---|---|
| | #Activation function | | Top 3 | | #Activation function | Top 5 | |
| | | | Training | Testing | | Training | Testing |
| | 1 activation function | Rbf_l1 (2000) | 0.9056 | 0.9017 | Rbf_linf (2000) | 0.9675 | 0.9656 |
| Malware | 2 activation function | Rbf_l1 (1000), Sigmoid (1000) | 0.9046 | 0.9018 | Rbf_l1 (1000), Sigmoid (1000) | 0.9707 | 0.9689 |
| | 3 activation function | Sigmoid (400), Tanh (1000), Rbf_linf(600) | 0.9037 | 0.8998 | Sigmoid (400), Tanh (1000), Rbf_linf(600) | 0.9668 | 0.9661 |

The Table 7 and 8 reports summarize information on the total number of evaluations in malware dataset which gained accuracy for HP-ELM in two experiments, with and without feature selection, which specifically focus on the number of activation functions. In the feature selection experiment, the total numbers of activation functions remained the same, but there was a significant difference in the accuracy of classification by increasing the number of feature selection. For instance, the accuracy of testing (0.9689) in the top 5 feature is higher than the accuracy of classification (0.9018) with the top 3 feature selection in the same activation function and the number of neurons. As shown in Table 8, in malware experiments with three activation functions, out of a total of 2000 neuron, the percentage of accuracy in top 5 feature was higher. In general, it can be seen that by increasing the number of activation functions without feature selection strategy, the percentage of accuracy is gradually dropping. On the other hand, by selecting more features and activation functions, the accuracy of classification increases moderately.

Table 9: Accuracy of HP-ELM without feature selection in CTU-13 dataset

| Dataset | #Activation function | Without feature selection | | |
|---|---|---|---|---|
| | | | Train | Test |
| CTU-13 | 1 activation function | Rbf_linf(2000) | 0.965 | 0.9625 |
| | 2 activation function | rbf_l1(1000), rbf_linf(1000) | 0.9642 | 0.9622 |
| | 3 activation function | Tanh (666), sigm(667), rbf_linf(667) | 0.9642 | 0.9616 |
| | 4 activation function | sigm(500), rbf_l1(500), rbf_l2(500), rbf_linf (500) | 0.9633 | 0.9611 |

Table 10: Accuracy of HP-ELM with feature selection in CTU-13 dataset

| | With feature selection | | | | | |
|---|---|---|---|---|---|---|
| | #Activation function | Top 3 | | #Act function | Top 5 | |
| | CTU-13 | Train | Test | | Train | Test |
| 1 act function | Rbf_linf(2000) | 0.9611 | 0.9592 | Rbf_linf (2000) | 0.9616 | 0.9589 |
| 2 act function | rbf_l1 (1000), rbf_linf(1000) | 0.9608 | 0.9588 | rbf_linf (1000), linear(1000) | 0.9605 | 0.9590 |
| 3 act function | Sigm(666), rbf_l1(667), rbf_linf(667) | 0.9607 | 0.9589 | rbf_l1(666), rbf_l12(667), rbf_linf(667) | 0.9615 | 0.9588 |
| 4 activation function | Tanh (500), sigm(500), rbf_l1(500), rbf_linf (500) | 0.9605 | 0.9582 | Tanh (500), sigm(500), rbf_l1(500), rbf_linf (500) | 0.9616 | 0.9585 |

Table 9 and 10 gives information on the accuracy of classification of HP-ELM for two experiments, with and without feature selection of CTU-13 dataset. In the feature selection experiment, HP-ELM was the highest accuracy leader with a low number of feature selection, and it is about 0.9592 for the top 3 features with one activation function.

## 8. Conclusion and future discussion

In this paper, we exploit HP-ELM to improve prediction stability for the training of (SLFNs). The presented model optimizes the input weights and hidden layers and provides more consistent performance in comparison to the other training models. Actual performance of the HP-ELM classifier tested under two real-world datasets, with various activation functions, neurons, layers, and different feature selections. The simulation results show that HP-ELM is a fast training method that reduces the root mean square error near to zero, it approaches accuracy ratio to one, and it achieves the feature optimization combination, and it also provides an excellent generalization performance on an SLFN and establishes a network intrusion detection model with the best overall performance. This finding, however, shows that despite promising results obtained by using the proposed model for this case study, it improved and further studies, which take more variables into account, will need to be undertaken.

**References**

1. Cerwall P, Jonsson P, Möller R, Bävertoft S, Carson S, Godor I, et al. Ericsson mobility report. On the Pulse of the Networked Society Hg v Ericsson. 2015.
2. Android Mobile Security Threats https://www.kaspersky.com/resource-center/threats/mobile2018.
3. Smith S. Cybercrime will Cost Businesses over $2 Trillion by 2019. Retrieved from Juniper Research: https://www. juniperresearch. com/press/pressreleases/cybercrime-cost-businesses-over-2trillion; 2015.
4. Report. Report: 2016 saw 8.5 million mobile malware attacks, ransomware and IoT threats on the rise. Available from: https://www.techrepublic.com/article/report-2016-saw-8-5-million-mobile-malware-attacks-ransomware-and-iot-threats-on-the-rise/.
5. Magdych JS, Rahmanovic T, McDonald JR, Tellier BE, Osborne AC, Herath NP. Secure gateway with firewall and intrusion detection capabilities. Google Patents; 2012.
6. Kolias C, Kambourakis G, Stavrou A, Voas J. DDoS in the IoT: Mirai and other botnets. Computer. 2017;50(7):80-4.
7. Kothmayr T, Hu W, Schmitt C, Bruenig M, Carle G, editors. Poster: Securing the internet of things with DTLS. Proceedings of the 9th ACM Conference on Embedded Networked Sensor Systems; 2011: ACM.
8. Enck W, Gilbert P, Han S, Tendulkar V, Chun B-G, Cox LP, et al. TaintDroid: an information-flow tracking system for realtime privacy monitoring on smartphones. ACM Transactions on Computer Systems (TOCS). 2014;32(2):5.
9. Wang T, Zhou J, Chen X, Wang G, Liu A, Liu Y. A Three-Layer Privacy Preserving Cloud Storage Scheme Based on Computational Intelligence in Fog Computing. IEEE Transactions on Emerging Topics in Computational Intelligence. 2018;2(1):3-12.
10. Altaher A. An improved Android malware detection scheme based on an evolving hybrid neuro-fuzzy classifier (EHNFC) and permission-based features. Neural Computing and Applications. 2017;28(12):4147-57.
11. Zhang Y, Pang J, Yue F, Cui J, editors. Fuzzy neural network for malware detect. Intelligent System Design and Engineering Application (ISDEA), 2010 International Conference on; 2010: IEEE.
12. Shalaginov A, Franke K. Automatic rule-mining for malware detection employing neuro-fuzzy approach. Norsk informasjonssikkerhetskonferanse (NISK). 2013;2013.
13. Tavallaee M, Stakhanova N, Ghorbani AA. Toward credible evaluation of anomaly-based intrusion-detection methods. IEEE Transactions on Systems, Man, and Cybernetics, Part C (Applications and Reviews). 2010;40(5):516-24.
14. Damopoulos D, Menesidou SA, Kambourakis G, Papadaki M, Clarke N, Gritzalis S. Evaluation of anomaly-based IDS for mobile devices using machine learning classifiers. Security and Communication Networks. 2012;5(1):3-14.
15. Huang G-B, Zhu Q-Y, Siew C-K, editors. Extreme learning machine: a new learning scheme of feedforward neural networks. Neural Networks, 2004 Proceedings 2004 IEEE International Joint Conference on; 2004: IEEE.
16. Huang G-B, Zhu Q-Y, Siew C-K. Extreme learning machine: theory and applications. Neurocomputing. 2006;70(1-3):489-501.
17. Huang G-B, Zhou H, Ding X, Zhang R. Extreme learning machine for regression and multiclass classification. IEEE Transactions on Systems, Man, and Cybernetics, Part B (Cybernetics). 2012;42(2):513-29.

18. Huang G-B. What are extreme learning machines? Filling the gap between Frank Rosenblatt's dream and John von Neumann's puzzle. Cognitive Computation. 2015;7(3):263-78.
19. Akusok A, Björk K-M, Miche Y, Lendasse A. High-performance extreme learning machines: a complete toolbox for big data applications. IEEE Access. 2015;3:1011-25.
20. Garcia S, Grill M, Stiborek J, Zunino A. An empirical comparison of botnet detection methods. computers & security. 2014;45:100-23.
21. Afifi F, Anuar NB, Shamshirband S, Choo K-KR. DyHAP: dynamic hybrid ANFIS-PSO approach for predicting mobile malware. PloS one. 2016;11(9):e0162627.
22. Saracino A, Sgandurra D, Dini G, Martinelli F. Madam: Effective and efficient behavior-based android malware detection and prevention. IEEE Transactions on Dependable and Secure Computing. 2018;15(1):83-97.
23. Egele M, Scholte T, Kirda E, Kruegel C. A survey on automated dynamic malware-analysis techniques and tools. ACM computing surveys (CSUR). 2012;44(2):6.
24. Burguera I, Zurutuza U, Nadjm-Tehrani S, editors. Crowdroid: behavior-based malware detection system for android. Proceedings of the 1st ACM workshop on Security and privacy in smartphones and mobile devices; 2011: ACM.
25. Narudin FA, Feizollah A, Anuar NB, Gani A. Evaluation of machine learning classifiers for mobile malware detection. Soft Computing. 2016;20(1):343-57.
26. Sun M, Li X, Lui JC, Ma RT, Liang Z. Monet: a user-oriented behavior-based malware variants detection system for android. IEEE Transactions on Information Forensics and Security. 2017;12(5):1103-12.
27. Salehi M, Amini M. Android Malware Detection using Markov Chain Model of Application Behaviors in Requesting System Services. arXiv preprint arXiv:171105731. 2017.
28. Poulter AJ, Johnson SJ, Cox SJ. Extensions and Enhancements to "the Secure Remote Update Protocol". Future Internet. 2017;9(4):59.
29. Huang G-B. An insight into extreme learning machines: random neurons, random features and kernels. Cognitive Computation. 2014;6(3):376-90.
30. Jiang X, Zhou Y, editors. Dissecting android malware: Characterization and evolution. 2012 IEEE Symposium on Security and Privacy; 2012: IEEE.
31. Shannon CE. A mathematical theory of communication. Bell system technical journal. 1948;27(3):379-423.
32. Ahmad Firdaus ZA. Mobile malware anomaly-based detection systems using static analysis features/Ahmad Firdaus Zainal Abidin: University of Malaya; 2017.
33. Kent JT. Information gain and a general measure of correlation. Biometrika. 1983;70(1):163-73.
34. Grill M, Nikolaev I, Valeros V, Rehak M, editors. Detecting DGA malware using NetFlow. 2015 IFIP/IEEE International Symposium on Integrated Network Management (IM); 2015: IEEE.
35. Roshan S, Miche Y, Akusok A, Lendasse A. Adaptive and online network intrusion detection system using clustering and Extreme Learning Machines. Journal of the Franklin Institute. 2018;355(4):1752-79.
36. Powers DM. Evaluation: from precision, recall and F-measure to ROC, informedness, markedness and correlation. 2011.
37. Gu Q, Li Z, Han J. Generalized fisher score for feature selection. arXiv preprint arXiv:12023725. 2012.
38. CTU. The CTU-13 dataset a labeled dataset with botnet-normal-and-background-traffic 2019 [cited 2019 27 Feb 2019]. Available from: https://mcfp.weebly.com/the-ctu-13-dataset-a-labeled-dataset-with-botnet-normal-and-background-traffic.html#.